\documentclass[times,sort&compress,3p]{elsarticle}
\journal{Journal of Multivariate Analysis}
\usepackage{amsmath,amsfonts,amssymb,amsthm}
\usepackage{booktabs,color,epsfig,graphicx,url,caption,subcaption}
\usepackage[colorlinks=true,citecolor=blue,urlcolor=blue,pdfstartview=FitB]{hyperref}

\usepackage{algorithm}
\usepackage{algpseudocode}

\newtheorem{theorem}{Theorem}[]

\newtheorem{corollary}{Corollary}[]

\newtheorem{lemma}{Lemma}[]

\DeclareMathOperator*{\argmin}{\arg\!\min}

\newcommand{\mbr}{\mathbb{R}}

\newcommand{\ms}{\boldsymbol}
\newcommand{\mb}{\mathbf}
\newcommand{\mc}{\mathcal}

\newcommand{\bop}{O_{\Pr}}
\newcommand{\sop}{o_{\Pr}}

\newcommand{\Var}{{\rm Var}}
\newcommand{\Cov}{{\rm Cov}}
\newcommand{\E}{{\rm E}}

\begin{document}

\begin{frontmatter}

\title{Predictor ranking and false discovery proportion control in high-dimensional regression}

\author[A1]{X. Jessie Jeng\corref{mycorrespondingauthor}}
\author[A2]{Xiongzhi Chen}

\address[A1]{Department of Statistics, North Carolina State University, Raleigh, NC 27695, USA}
\address[A2]{Department of Mathematics and Statistics, Washington State University, Pullman, WA 99164, USA}

\cortext[mycorrespondingauthor]{Corresponding author. Email address: \url{xjjeng@ncsu.edu}}

\begin{abstract}
We propose a ranking and selection procedure to prioritize relevant predictors and control false discovery proportion (FDP) of variable selection.
Our procedure utilizes a new ranking method built upon the de-sparsified Lasso estimator. We show that the new ranking method achieves the optimal order of minimum non-zero effects in ranking relevant predictors ahead of irrelevant ones.
Adopting the new ranking method,
we develop a variable selection procedure to asymptotically control FDP at a  user-specified level.
We show that our procedure can consistently estimate the FDP of variable selection as long as the de-sparsified Lasso estimator is asymptotically normal.
In numerical analyses, our procedure compares favorably to existing methods in ranking efficiency and FDP control when the regression model is relatively sparse.
\end{abstract}


\end{frontmatter}

\section{Introduction}

In the past fifteen years, impressive progress has been made in high-dimensional statistics where the number of unknown parameters can greatly exceed the sample size.
%More recent studies describe the uncertainty in estimation by constructing confidence intervals and assigning $p$-values; see a nice summary in  \cite{hdi2015}.
We consider a sparse linear model
\begin{equation*}\label{def:model}
y= \mb{x}^\top \ms{\beta} +\varepsilon,
\end{equation*}
where $y$ is the response variable, $\mb{x} = (x_1, \ldots, x_p)^\top$ the vector of predictors, $\ms{\beta} = (\beta_1, \ldots, \beta_p)^\top$ the unknown coefficient vector, and $\varepsilon$ the random error.
Our goal is to simultaneously test
\[
H_{0j}: \beta_j = 0 \quad \text{against} \quad H_{1j}: \beta_j \ne 0 \quad \text{for} \qquad j = 1, \ldots, p
\]
and select a predictor $X_j$ into the model if $H_{0j}$ is rejected.

Much work has been conducted on point estimation of $\ms{\beta}$; see, for instance, Chapters 1-10 of \cite{BVdGeer11}. Among the most popular point estimators, Lasso benefits from the geometry of the $L_1$ norm penalty to shrink some coefficients exactly to zero and hence performs variable selection \citep{Tibshirani:1996}.
The Lasso estimator $\hat{\ms{\beta}}$ possesses desirable properties including the oracle inequalities on $\Vert \hat{\ms{\beta}} - \ms{\beta} \Vert_q$ for $q \in [1, 2]$ \citep{Bickel2009, BVdGeer11}.
However, it is difficult to characterize the distribution of the Lasso estimator and assess the significance of selected variables.

Recently, the focus of research in high-dimensional regression has been shifted to confidence intervals and hypothesis testing for $\ms{\beta}$. Substantial progress has been made in
\cite{CaiGuo17} \cite{dezeure2017high}, \cite{JM2018}, \cite{Lee16}, \cite{Lockhart14}, \cite{MLB09}, \cite{vandegeer2014}, \cite{Wasserman09}, \cite{Zhang:2014}, etc.
In particular, innovative methods have been developed to enable multiple hypothesis testing on $\ms{\beta}$. For example, \cite{Buhlmann2013} and \cite{ZhangX:2016} propose to control  family-wise error rate (FWER) under the dependence imposed by $\ms{\beta}$ estimation. Methods to control false discovery rate (FDR, \cite{Benjamini:1995}) have been developed in \cite{Barber15}, \cite{Bogdan15}, \cite{CandesPanning}, \cite{Gsell16}, \cite{JiZhao14}, \cite{Su16}, etc.

In this paper, we aim to prioritize relevant predictors in predictor ranking and select variables by controlling
false discovery proportion (FDP, \cite{Genovese:2002}).
FDP is the ratio of the number of false positives to the number of total rejections. Given an experiment, FDP is realized but unknown.
%FDP is arguably more relevant than FDR, which is the expected value of FDP, because FDP is directly related to the current experiment \citep{Fan2012fdp}.
In the literature of multiple testing, estimating FDP under dependence has been studied in, e.g., \cite{efron2007correlation}, \cite{Fan2012fdp} and \cite{Friguet09}.

We propose the DLasso-FDP procedure, which ranks and selects predictors in linear regression based on the de-sparsified Lasso (DLasso) estimator and its limiting distribution \citep{Zhang:2014, vandegeer2014}.
We show that ranking the predictors by the standardized DLasso estimator achieves the optimal order of the minimum non-zero effect for ranking relevant predictors ahead of irrelevant ones when the dimension $p$, sample size $n$, and the number of non-zero coefficients $s_0$ satisfy $s_0 = o(n/\ln p)$.
Further, we develop consistent estimators of the FDP and marginal FDR for variable selection based on the standardized DLasso estimator.
Unlike in conventional studies on FDP and FDR where the null distributions of test statistics are exact, the null distribution of the DLasso estimator can only be approximated asymptotically, and the approximation errors for all estimated regression coefficients need to be considered conjointly to estimate FDP.
Our simulation studies support our theoretical findings and demonstrate that DLasso-FDP
compares favorably with existing methods in ranking efficiency and FDP control, especially when the regression model is relatively sparse.

%Specifically, we develop the DLasso-FDP procedure, which ranks the predictors by the standardized DLasso estimates and selects the top-ranked predictors by controlling the FDP.

The rest of the article is organized as follows. \autoref{secTheory} provides theoretical analyses on the ranking efficiency of the standardized DLasso estimator and consistent estimation of the FDP and marginal FDR of the DLasso-FDP procedure. Numerical analyses are presented in \autoref{sec:Simulation}.
%and an application of the DLasso-FDP procedure to a genetic association study is demonstrated in \autoref{sec:Application}.
\autoref{sec:disc} provides further discussions. All proofs are presented in the appendix.

\section{Theory and Method\label{secTheory}}

\subsection{Notations}

We collect notations that will be used throughout the article.
The symbols $O (\cdot)$ and $o(\cdot)$ respectively denote Landau's big O and small o notations, for which accordingly $\bop (\cdot)$ and $\sop (\cdot)$ their probabilistic versions.
%For two positive sequences $\{a_m\}_{m \geq 1}$ and $\{a^{\prime}_m\}_{m \geq 1}$, $a_m \ll a^{\prime}_m$ means that $\lim_{m \to \infty}a_m/a^{\prime}_m = 0$, and $a_m \gg a^{\prime}_m$ means that $\lim_{m \to \infty}a^{\prime}_m/a_m = 0$.
The symbol $C$ denotes a generic, finite constant whose values can be different at different occurrences.

For a matrix $\mb{M}$, $\mb{M}_{ij}$ denotes its $\left(  i,j\right)  $ entry,
the $q$-norm $\left\Vert \mb{M}\right\Vert _{q}=\left(  \sum_{i,j}\left\vert
\mb{M}_{ij}\right\vert ^{q}\right)  ^{1/q}$ for $q>0$, $\infty$-norm $\left\Vert \mb{M}\right\Vert _{\infty
}=\max_{i,j}\left\vert \mb{M}_{ij}\right\vert $, and $\left\Vert \mathbf{M}\right\Vert _{1,\infty}$ is the maximum of the
$1$-norm of each row of $\mathbf{M}$.
If $\mb{M}$ is symmetric, $\sigma_i(\mb{M})$ denotes the $i$th smallest eigenvalue of $\mb{M}$.
The symbol $\mb{I}$ denotes the identity matrix.
A vector $\mb{v}$ is always a column vector whose $i$th component is denoted by $v_i$.
For a set $A$, $\vert A \vert$ denotes its cardinality and $1_A$ its indicator. $a \vee b = \max\{a,b\}$ for two real numbers $a$ and $b$.

\subsection{Regression model and the de-sparsified Lasso estimator}

Given $n$ observations from the model $y= \mb{x}^\top \ms{\beta}  +\varepsilon$, we have
\begin{equation}\label{def:model}
\mb{y}=\mb{X}\ms{\beta}+\ms{\varepsilon},
\end{equation}
where $\mb{y} = (y_1, \ldots, y_n)^\top$ and $\mb{X} = [\mb{x}_1,\ldots,\mb{x}_p]\in\mathbb{R}^{n\times p}$. We assume $\ms{\varepsilon}\sim\mathcal{N}_{n}\left(  0,\sigma^{2}\mb{I}\right)$ and $\sigma^{2}=O\left(  1\right)$ in this work.
Let $S_{0}=\left\{j:\beta_{j}\neq0\right\}  $ and $s_{0}=\left\vert S_{0}\right\vert $.
The Lasso estimator is
\begin{equation} \label{eqLasso}
\hat{\ms{\beta}} = \hat{\ms{\beta}}(\lambda) = \arg\min_{\ms{\beta} \in \mathbb{R}^p} (\Vert \mb{y} - \mb{X} \ms{\beta} \Vert_2^2/n + 2\lambda \Vert \ms{\beta}\Vert_1).
\end{equation}
Let $\hat{\ms{\Sigma}}=n^{-1}\mb{X}^\top\mb{X}$. To obtain the de-sparsified Lasso estimator for $\ms{\beta}$ as in \cite{vandegeer2014} and \cite{Zhang:2014}, a matrix $\hat{\ms{\Theta}} \in \mbr^{p \times p}$ such
that $\hat{\ms{\Theta}}\hat{\ms{\Sigma}}$ is close to $\mb{I}$ is obtained by Lasso for nodewise
regression on $\mb{X}$ as in \cite{MB06}.
Let $\mb{X}_{-j}$ denote the matrix obtained by removing the $j$th column of $\mb{X}$.
For each $j=1,\ldots,p$, let%
\begin{equation}
\hat{\ms{\gamma}}_{j}=\argmin_{\ms{\gamma}\in\mathbb{R}^{p-1}}\left(  n^{-1}\left\Vert
\mb{x}_j-\mb{X}_{-j}\ms{\gamma}\right\Vert _{2}^{2}+2\lambda_{j}\left\Vert \ms{\gamma}
\right\Vert _{1}\right)  \label{eqNodewiseReg}%
\end{equation}
with components $\hat{\gamma}_{j,k},k=1,\ldots,p$ and $k\neq j$. Further, define%
\begin{equation*}%\label{eqTauhat}
\hat{\tau}_{j}^{2}=  n^{-1}%
\left\Vert \mb{x}_j-\mb{X}_{-j}\hat{\ms{\gamma}}_{j}\right\Vert _{2}^{2}+2\lambda
_{j}\left\Vert \hat{\ms{\gamma}}_{j}\right\Vert _{1}
\end{equation*}
and
\begin{equation}
\hat{\ms{\Theta}}=\operatorname{diag}\left(  \hat{\tau}_{1}^{-2},\cdots,\hat{\tau
}_{p}^{-2}\right)  \left(
\begin{array}
[c]{cccc}%
1 & -\hat{\gamma}_{1,2} & \cdots & -\hat{\gamma}_{1,p}\\
-\hat{\gamma}_{2,1} & 1 & \cdots & -\hat{\gamma}_{2,p}\\
\vdots & \vdots & \vdots & \vdots\\
-\hat{\gamma}_{p,1} & -\hat{\gamma}_{p,2} & \cdots & 1
\end{array}
\right) \nonumber.  %\label{Thetahat}%
\end{equation}
The estimator
\begin{equation}\label{eqRelaxedEst}
\hat{\mb{b}} = \hat{\ms{\beta}}+n^{-1}%
\hat{\ms{\Theta}}\mb{X}^\top (  \mb{y}-\mb{X}\hat{\ms{\beta}} )
\end{equation}
is referred to as the de-sparsified Lasso  (DLasso) estimator. This implies
\begin{equation*} %\label{eq:b_decomp}
\sqrt{n} (\hat{\mb{b}}-\ms{\beta} )
=n^{-1/2}\hat{\ms{\Theta}}%
\mb{X}^\top\ms{\varepsilon}-\ms{\delta} = \mb{w} - \ms{\delta},
\end{equation*}
where%
\[
\mb{w}|\mb{X}\sim\mathcal{N}_{p} (  0,\sigma^2 \hat{\ms{\Omega}}), \qquad \hat{\ms{\Omega}} = \hat{\ms{\Theta}} \hat{\ms{\Sigma}}\hat{\ms{\Theta}}^\top,
\]
and
\begin{equation*}%\label{eqDelta}
\ms{\delta}=\sqrt{n} (  \hat{\ms{\Theta}}\hat{\ms{\Sigma}}-\mb{I} )   (  \hat{\ms{\beta}}-\ms{\beta} ).
\end{equation*}

Since the distribution of $\mb{w}|\mb{X}$ is fully specified, it is essential to study $\ms{\delta}$ to derive the distribution of $\hat{\mb{b}}$. We adopt the result in  \cite{JM2018}, which provides an explicit bound on the magnitude of $\ms{\delta}$.
%identifying the distribution of $\hat{\mb{b}}$ requires an accurate estimate on the magnitude of $\ms{\delta}$. Among the estimates of $\ms{\delta}$ given in \cite{Zhang:2014}, \cite{vandegeer2014} and  \cite{Javanmard2016}, the one in \cite{Javanmard2016}, which we will adopt in this work and restate next, is probably the best.
Let $\ms{\Theta}=\ms{\Sigma}^{-1}$, $s_{j}=\left\vert\left\{  k\neq j:\ms{\Theta}_{jk}\neq0\right\}\right\vert$ and $s_{\max}=\max_{1\le j \le p}s_{j}$. Note that $s_j$ can be regarded as the number of non-zero coefficients when regressing $X_j$ on the remaining predictors.
Suppose the following hold:
\begin{description}

\item[A1)] Gaussian random design: the rows of $\mb{X}$ are i.i.d. $\mathcal{N}_{p}\left(
0,\ms{\Sigma}\right)  $ for which $\ms{\Sigma}$ satisfies:

\begin{description}
  \item[A1a)] $\max_{1\leq j\leq p}\boldsymbol{\Sigma}_{jj}\leq1$.

\item[A1b)] $0<C_{\min}\leq\sigma_{1}\left(  \boldsymbol{\Sigma}\right)
\leq\sigma_{p}\left(  \boldsymbol{\Sigma}\right)  \leq C_{\max}<\infty$ for
constants $C_{\min}$ and $C_{\max}.$

\item[A1c)]  $\rho\left(  \Sigma,C_{0}s_{0}\right)  \leq\rho$ for some
constant $\rho>0$, where $C_{0}=32C_{\max}C_{\min}^{-1}+1$,
\[
\rho\left(  \mathbf{A},k\right)  =\max_{T\subseteq\left[  p\right]
,\left\vert T\right\vert \leq k} \big\Vert \left(  \mathbf{A}_{T,T}\right)
^{-1} \big\Vert _{1,\infty}
\]
for a square matrix $\mb{A}$,
$\left[  p\right]  =\left\{  1,...,p\right\}  $, $\mathbf{A}_{T,T}$ is a
submatrix formed by taking entries of $\mathbf{A}$ whose row and column
indices respectively form the same subset $T$.
\end{description}

\item[A2)] Tuning parameters:
for the Lasso in \eqref{eqLasso}, $\lambda= 8\sigma\sqrt{n^{-1}\ln
p}$; %for some $\kappa\in\left[  8,\kappa_{\max}\right]  $ and some finite positive constant $\kappa_{\max} \ge 8$;
for nodewise regression in \eqref{eqNodewiseReg},  $\lambda_{j}=\tilde{\kappa}\sqrt{n^{-1}\ln p}, j=1,\ldots,p$ for a suitably
large universal constant $\tilde{\kappa}$.

\end{description}
We rephrase Theorem 3.13 of \cite{JM2018} for unknown $\boldsymbol{\Sigma}$ as follows.

\begin{lemma}\label{lmJM16}
Consider model (\ref{def:model}). Assume A1) and A2).
Then there exist positive constants $c$ and $c^{\prime}$ depending only on $C_{\min}$, $C_{\max}$ and $\tilde{\kappa}$
such that, for $\max\{s_{0},s_{\max}\} < c n /\ln p$, the probability that%
\begin{equation*}
\left\Vert \boldsymbol{\delta}\right\Vert _{\infty} \le c^{\prime}\rho\sigma\sqrt
{\frac{s_{0}}{n}}\ln p+c^{\prime}\sigma\min\left\{  s_{0},s_{\max}\right\}  \frac{\ln
p}{\sqrt{n}}%\label{eqJM16}%
\end{equation*}
is at least  $1-2pe^{-16^{-1}ns_{0}^{-1}C_{\min}}-pe^{-cn}-6p^{-2}$.
\end{lemma}

Lemma \ref{lmJM16} provides an explicit bound on the magnitude of $\ms{\delta}$, and hence the difference between the distribution of the DLasso estimator $\hat{\mb{b}}$ and the normally distributed variable $\mb{w}|\mb{X}$. This is very helpful for our subsequent studies.
%on ranking consistency and variable selection with FDP control based on the standardized DLasso estimate.

\subsection{Ranking efficiency of DLasso estimator}

In general, variable selection procedures often rank predictors by some measure of importance and select a subset of top-ranked predictors based on a selection criterion. For instance, the Lasso ranks predictors by the Lasso solution path and selects a subset of top-ranked predictors by, for example, cross validation.
In this paper, we propose to rank the predictors by the standardized DLasso estimator and select the top-ranked predictors via FDP control.
The standardized DLasso estimator is constructed as
\begin{equation} \label{def:Z}
z_j = \sqrt{n} \hat b_j \sigma^{-1} \hat{\ms{\Omega}}_{jj}^{-1/2}, \qquad 1 \le j \le p.
\end{equation}
We rank the predictors by their absolute values of $z_j$  in a decreasing order. Let
$I_{0}=\left\{  1\leq j\leq p:\beta_{j}=0\right\}$ and $p_0 = |I_0|$. We say that all relevant predictors are asymptotically ranked ahead of any irrelevant predictor if
\begin{equation*} %\label{eq12}%
\lim_{p\rightarrow\infty}\Pr\left( \min_{j\in S_{0}}\left\vert
z_{j}\right\vert > \max_{j\in I_{0}}\left\vert z_{j}\right\vert \right) =1.
\end{equation*}
Note that although  the DLasso estimates are asymptotically normally distributed given $\mb{X}$, their asymptotic covariance matrix
$\sigma^2 \hat{\ms{\Omega}}$ ($\hat{\ms{\Omega}} =  \hat{\ms{\Theta}} \hat{\ms{\Sigma}}\hat{\ms{\Theta}}^\top$) is not a sparse matrix.  The following theorem provides insights for the efficiency of ranking predictors by $|z_j|$ under such covariance dependence.

\begin{theorem} \label{ThmSeparation}
Consider model (\ref{def:model}) and the standardized DLasso estimator  $\left\{
z_{j}\right\}_{j=1}^{p}$ in (\ref{def:Z}). Let
\[
C_{p}=\ln (p^{2} / 2\pi)+\ln\ln (p^{2} / 2\pi)
\]
and
\[
B_{p}\left(  s_{0},n,\boldsymbol{\Sigma}\right)  =c^{\prime}\rho\sigma
	\sqrt{\frac{s_{0}}{n}}\ln p+c^{\prime}\sigma\min\left\{  s_{0},s_{\max}\right\}
	\frac{\ln p}{\sqrt{n}}.
\]
Assume A1) and A2). If $s_0 \le p_0$, $\max\{s_0, s_{\max}\} = o(n/\ln p)$ and
\begin{equation} \label{eq22pA}
	\beta_{\min} := \min_{j\in S_{0}} \left\vert \beta_{j}\right\vert \geq 2  n^{-1/2} \left\{ \sqrt{C_{\min}^{-1}C_{\max}}B_{p}\left(  s_{0},n,\boldsymbol{\Sigma}\right) + \sigma \sqrt{C_{\max}} (1+a)\sqrt{C_{p_{0}}}\right\}
\end{equation}
for some constant $a>0$,
then the standardized DLasso estimator asymptotically rank all relevant predictors ahead of any irrelevant ones, i.e., $\Pr\left( \min_{j\in S_{0}}\left\vert z_{j}\right\vert > \max_{j\in I_{0}}\left\vert z_{j}\right\vert \right) \to 1$  as $s_0 \to \infty$.
\end{theorem}

Condition (\ref{eq22pA}) on $\beta_{\min}$ is imposed to separate relevant predictors from irrelevant ones. Note that condition (\ref{eq22pA}) implies $\beta_{\min} > C \sqrt{\ln p / n}$, and the order of $\sqrt{\ln p / n}$ is optimal for perfect separation of signals from noise. In other words, under suitable conditions, ranking variables by $\left\{
|z_{j}|\right\}_{j=1}^{p}$ obtains the optimal order of $\beta_{\min}$ for  perfect separation.
%Such optimal property is also achieved by Lasso with $s_{0} = o(n/\ln p)$.
Further, compared to \autoref{lmJM16}, the stronger condition in \autoref{ThmSeparation} on $s_{\max}$, i.e., $s_{\max} = o(n/\ln p)$, ensures $\Vert\hat{\ms{\Omega}} -\ms{\Sigma}^{-1}\Vert_{\infty} =\sop\left(1\right)$, so that the standardization of each $\hat{b}_j$ in (\ref{def:Z}) is proper.

\subsection{Consistent estimation of FDP and marginal FDR}

Recall that we are simultaneously testing $H_{0j}: \beta_j = 0$ versus $H_{1j}: \beta_j \ne 0$ for $j = 1, \ldots, p$ and
selecting predictor $X_j$ into the model whenever $H_{0j}$ is rejected.
The findings on the ranking efficiency of the standardized DLasso help us develop a variable selection procedure with the following rejection rule:
\begin{equation}\label{rejectionRule}
\text{reject } H_{0j} \text{ whenever } |z_j| > t \text{\ for a fixed rejection threshold } t>0.
\end{equation}
Define
$R_{\mb{z}}\left(  t\right)  =\sum_{j=1}^{p}1_{\left\{  \left\vert z_{j}\right\vert >t\right\}}$ as the number of discoveries and $V_{\mb{z}}\left(  t\right) =\sum_{j\in I_{0}} 1_{\left\{  \left\vert z_{j}\right\vert >t\right\}  }$ the number of false discoveries.
Then the FDP of the procedure at rejection threshold $t$ is
\[
FDP_{\mb{z}}\left(  t\right)  =\frac{V_{\mb{z}}\left(  t\right)  }{R_{\mb{z}}\left(  t\right) \vee 1 }.
\]

To control the FDP of the procedure at a prespecified level, we propose to consistently estimate $FDP_{\mb{z}}\left(  t\right)$ for any fixed $t$. To this end, we state an extra assumption:
\begin{description}
  \item [A3)] Sparsities of $\ms{\beta}$ and $\ms{\Theta}$: $\max\{s_{0},s_{\max}\} = o(n /\ln p)$,  $\min\{s_{\max},s_0\} = o(\sqrt{n}/\ln p)$, $s_0 = o \left(n / (\ln p)^2\right)$ and $s_0 = o(p)$.
\end{description}

\noindent Assumption A3), together with \autoref{lmJM16}, ensures $\Vert \ms{\delta} \Vert_{\infty}=\sop(1)$ \citep{JM2018}. This is sufficient for us to construct a consistent estimator of $FDP_{\mb{z}}\left(  t\right)$, i.e.,
\begin{equation*}%\label{fdpEst}
\widehat{FDP}(t) = \frac{2 p \Phi(-t)}{R_{\mb{z}}\left(t\right) \vee 1},
\end{equation*}
where $\Phi$ is the cumulative distribution function (CDF) of the standard normal random variable. Note that $\widehat{FDP}(t)$ is observable based on $\left\{z_{j}\right\}_{j=1}^{p}$, and $\left\{z_{j}\right\}_{j=1}^{p}$ are dependent with non-sparse covariance matrix.
\begin{theorem}\label{ThmSLLNFDP}
Consider model (\ref{def:model}) and the standardized DLasso estimator $\left\{
z_{j}\right\}_{j=1}^{p}$ in (\ref{def:Z}).
Assume A1) to A3). Then
\begin{equation}\label{eqSLLNfdpA}
\widehat{FDP}(t) -  FDP_\mb{z}\left(  t\right)
 =\sop(1).
\end{equation}
\end{theorem}

\autoref{ThmSLLNFDP} shows that $FDP_\mb{z}\left(  t\right)$ can be consistently estimated by the observable quantity $\widehat{FDP}(t)$ when $\ms{\beta}$ and $\ms{\Theta}$ are sparse in the sense of assumption A3). Moreover, no additional assumptions other than those to ensure asymptotic normality of the DLasso estimator are needed when $\mb{X}$ is from Gaussian random design.
%[xc: the assumption on $\ms{\Theta}$ is a restriction on $\mb{X}$]

An analogous result can be obtained for estimating the marginal FDR, which is defined as
\[
mFDR_{\mb{z}}\left(  t\right)  =\frac{\E\{V_{\mb{z}}\left(  t\right)\}  }{	\E\{R_{\mb{z}}\left(  t\right) \vee 1 \}  }.
\]
Marginal FDR was proposed in \cite{Sun:2007} and has been proved to be close to FDR when test statistics are independent. Here, we have:

\begin{corollary}\label{Thm:mFDR}
Under the conditions in Theorem \ref{ThmSLLNFDP},
\begin{equation}\label{eqSLLNfdpB}
\widehat{FDP}(t) - mFDR_{\mb{z}}\left(  t\right)  =\sop(1).
\end{equation}
\end{corollary}

\subsection{Algorithm for the DLasso-FDP procedure}
%{Variable selection with FDP control: DLasso-FDP procedure}

Once we are able to consistently estimate the FDP of the procedure defined by \eqref{rejectionRule}, for a user-specified $\alpha \in (0,1)$ we can determine the rejection threshold $t_{\alpha}$ such that $\widehat{FDP}_{\mb{z}}\left( t_{\alpha}\right) \le \alpha$ and then reject $H_{0j}$ if $|z_j| > t_\alpha$ for each $j$.
This procedure, which we call the De-sparsified Lasso FDP (DLasso-FDP) procedure, will have its FDP asymptotically bounded by $\alpha$. The implementation of the procedure is provided in Algorithm 1.

%%%%%
\begin{algorithm}[H]
\begin{algorithmic}[1]
%\small
\caption{DLasso-FDP}\label{alg}

\State \parbox[t]{\dimexpr\linewidth-\algorithmicindent}{
Calculate the DLasso estimator by (\ref{eqRelaxedEst}) and obtain $\left\{z_{j}\right\}_{j=1}^{p}$ by (\ref{def:Z}).
\strut}

\State \parbox[t]{\dimexpr\linewidth-\algorithmicindent}{
Rank the predictors by the absolute values of $\left\{z_{j}\right\}_{j=1}^{p}$ so that $|z_{(1)}| > \ldots > |z_{(p)}|$.
\strut}

\State \parbox[t]{\dimexpr\linewidth-\algorithmicindent}{
Specify an $\alpha \in (0,1)$ for FDP control; e.g., $\alpha =0.1$.
\strut}

\State \parbox[t]{\dimexpr\linewidth-\algorithmicindent}{
Find the minimum value of $t$, denoted by $t_\alpha$, such that $\widehat{FDP}(t) \le \alpha$.
\strut}

\State \parbox[t]{\dimexpr\linewidth-\algorithmicindent}{
Select the top-ranked predictors with $|z_{(j)}|>t_\alpha$.
\strut}

\end{algorithmic}
\end{algorithm}

The following corollary summarizes  the asymptotic control of FDP and mFDR by the DLasso-FDP procedure.
%Its proof is straightforward from Theorem \ref{ThmSLLNFDP} and Corollary \ref{Thm:mFDR}, and thus omitted.
\begin{corollary}\label{Cor:FDPcontrol}
	Given a fixed $\alpha \in (0, 1)$, select predictors by  the DLasso-FDP procedure described in Algorithm \ref{alg}. Then, under the conditions in Theorem \ref{ThmSLLNFDP},
	\[
	\Pr\left\{  FDP_{\mathbf{z}}\left(  t_{\alpha}\right)  \leq\alpha\right\}
	\rightarrow 1 \qquad \text{and} \qquad
	\Pr\left\{  mFDR_{\mathbf{z}}\left(
	t_{\alpha}\right)  \leq\alpha\right\}  \rightarrow1.
	\]
\end{corollary}

\section{Numerical Analysis\label{sec:Simulation}}

In the following examples, the linear model (\ref{def:model}) is simulated with $p=200$, $\ms{\varepsilon} \sim \mc{N}_{n}(0,\mb{I})$, and each row of $\mb{X} \sim \mc{N}_p(0, \ms{\Sigma})$. We use the Erg{\"o}s-R{\'e}nyi random graph in \cite{CLZ16} to generate the precision matrix
$\ms{\Theta} = \ms{\Sigma}^{-1}$ with $s_{\max}$ generated from the binomial distribution $\mathcal{B}(p, 0.05)$, such that
the nonzero elements of $\ms{\Theta}$ are  randomly located in each of its rows with magnitudes randomly generated from the uniform distribution $\mathcal{U}[0.4, 0.8]$. Without loss of generality, $\beta_j, j=1,\ldots, s_0$, are nonzero coefficients with the same value. We consider settings of different sample size ($n$), number of nonzero coefficients ($s_0$), and effect size of $\beta_1, \ldots \beta_{s_0}$. We obtain the DLasso estimates using the \textsf{R} package \texttt{hdi} and  derive $\mb{z}$ by (\ref{def:Z}).

\noindent {\bf Example 1: Ranking efficiency based on DLasso estimate.}
We compare the ranking of $\left\{|z_{j}|\right\}_{j=1}^{p}$ with the ranking based on Lasso solution path, which is generated by the \textsf{R} package \texttt{glmnet}. The efficiency of ranking is illustrated using the FDP-TPP curve, where TPP represents true positive proportion and is defined as the number of true positives divided by $s_0$.

\begin{figure}[!h]
	%\centering
	
	%\vspace{-0.15in}
	\begin{subfigure}[b]{0.45\textwidth}
		\includegraphics[width=\textwidth, height=0.3\textheight]{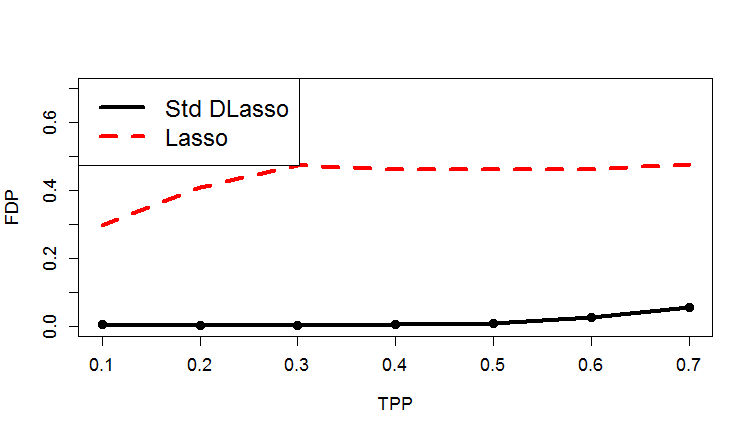}
		\caption{$s_0=10$, $n=100$, $\beta_{1} = 0.5$.}
		\label{fig:rank_1}
	\end{subfigure}
	\hfill
	\begin{subfigure}[b]{0.45\textwidth}
		\includegraphics[width=\textwidth, height=0.3\textheight]{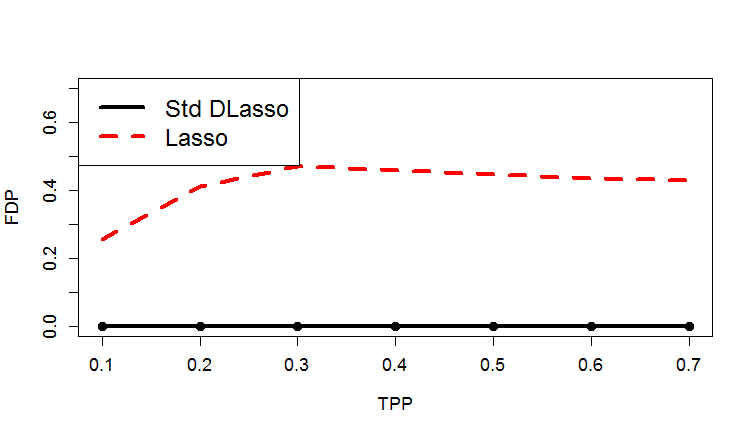}
		\caption{$s_0=10$, $n=100$, $\beta_{1} = 1$.}
		\label{fig:rank_2}
	\end{subfigure}
	\hfill
	\begin{subfigure}[b]{0.45\textwidth}
		\includegraphics[width=\textwidth, height=0.3\textheight]{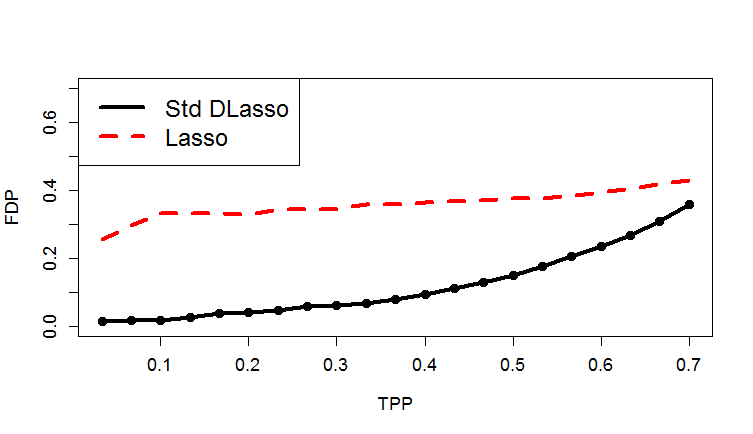}
		\caption{$s_0=30$, $n=150$, $\beta_{1} = 0.5$.}
		\label{fig:rank_3}
	\end{subfigure}
	\hfill
	\begin{subfigure}[b]{0.45\textwidth}
	\includegraphics[width=\textwidth, height=0.3\textheight]{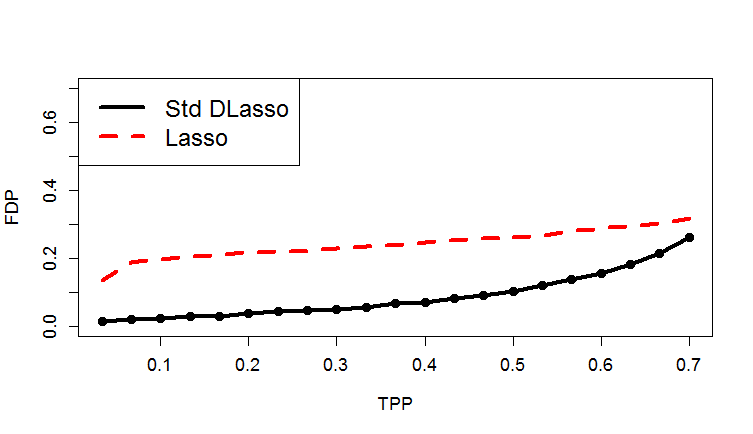}
	\caption{$s_0=30$, $n=150$, $\beta_{1} = 1$. }
	\label{fig:rank_4}
	\end{subfigure}
\caption{Comparison in ranking efficiency of the standardized DLasso estimate (solid line) and Lasso solution path (dashed line). } \label{fig:ranking}
\end{figure}

For a given TPP $\in \{ 1/s_0, \ldots, s_0/s_0\}$, we measure the corresponding FDP, which is the  price to pay in false positives for retaining the given TPP level. Consequently, a more efficient method for ranking would have a lower FDP-TPP curve.
Figure \ref{fig:ranking} reports the mean values of the FDP-TPP curves over 100 replications for different methods.
It shows that the ranking of $\left\{|z_{j}|\right\}_{j=1}^{p}$  is more efficient than that based on the Lasso solution path in prioritizing relevant predictors over irrelevant ones under finite sample. The reason, we think, is because DLasso mitigates the bias induced by Lasso shrinkage.

\begin{figure}[!h]
	%\centering
	
	\begin{subfigure}[b]{0.45\textwidth}
		\includegraphics[width=\textwidth, height=0.28\textheight]{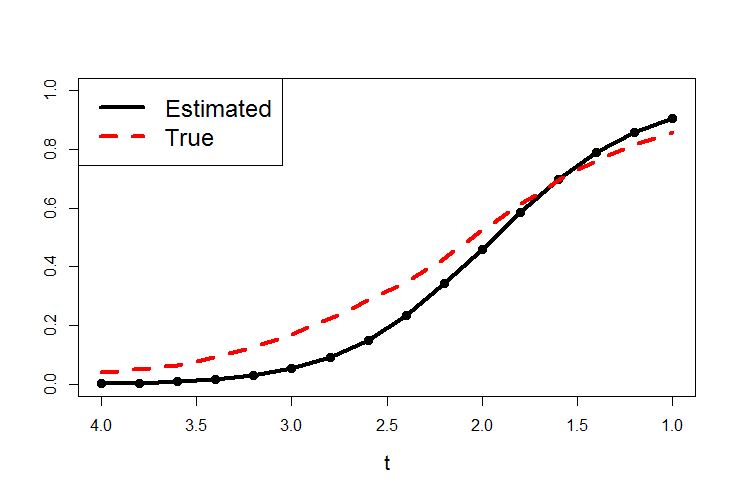}
		\caption{$s_0 = 10$,  $n=100$}
		\label{fig:f1}
	\end{subfigure}
	\hfill
	\begin{subfigure}[b]{0.45\textwidth}
		\includegraphics[width=\textwidth, height=0.28\textheight]{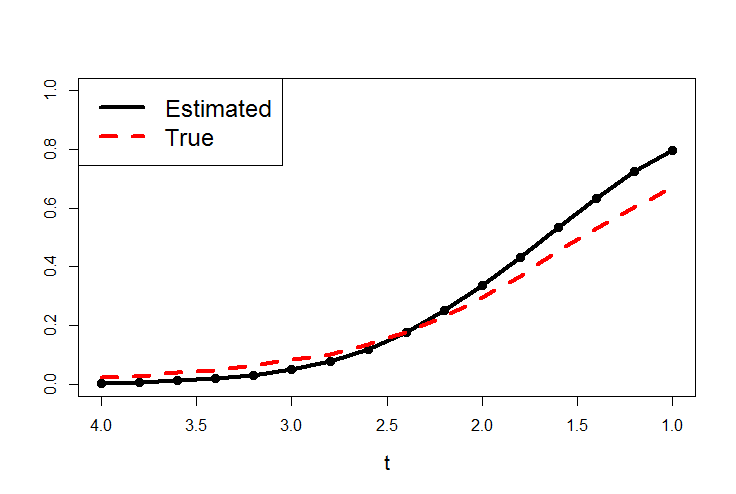}
		\caption{$s_0=10$, $n=150$}
		\label{fig:f2}
	\end{subfigure}
	\begin{subfigure}[b]{0.45\textwidth}
		\includegraphics[width=\textwidth, height=0.28\textheight]{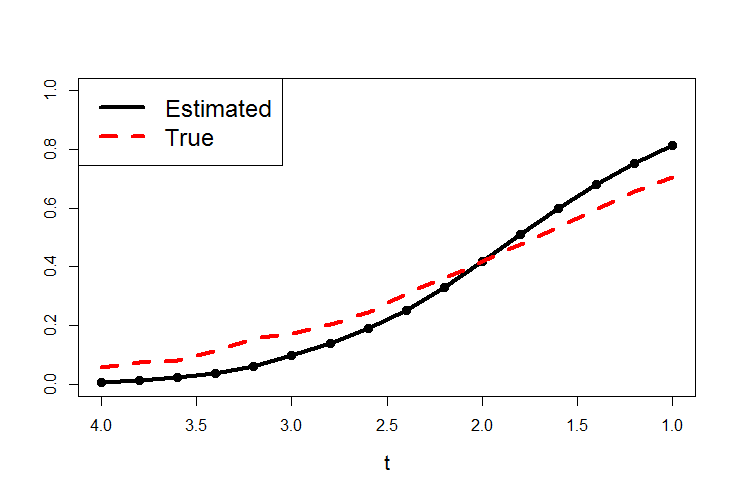}
		\caption{$s_0=30$, $n=100$}
		\label{fig:f3}
	\end{subfigure}
	\hfill
	\begin{subfigure}[b]{0.45\textwidth}
		\includegraphics[width=\textwidth, height=0.28\textheight]{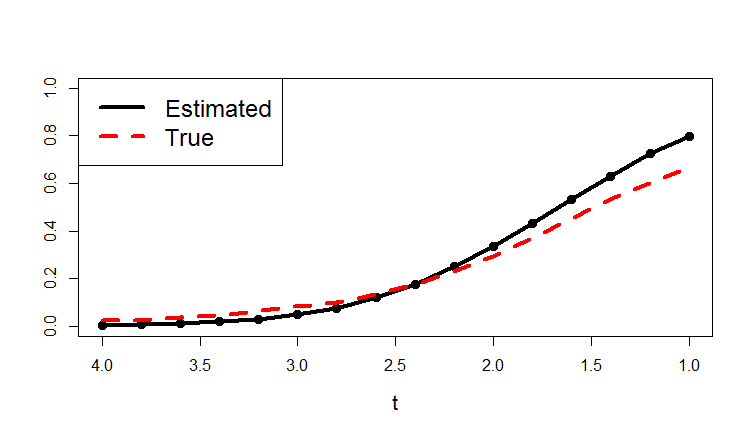}
		\caption{$s_0=30$, $n=150$}
		\label{fig:f4}
	\end{subfigure}
\caption{Mean values of the true FDP (dashed line) and estimated FDP (solid line) with $p=200$ and $\beta_1=0.5$. } \label{fig:FDR}
\end{figure}

\noindent {\bf Example 2: Estimation of FDP}.
In this example, we compare our estimated FDP with the true FDP in the settings with $p=200$, $\beta_1=0.5$, $n=100$ or $150$, and $s_0=10$ or 30.  Figure \ref{fig:FDR} presents the  empirical mean of our estimated FDP and the empirical mean of the true FDP for different $t$ values. It can be seen that (i) the mean values of the two statistics generally agree with each other in all cases, (ii) the estimated FDP tends to be lower than true FDP for larger $t$ values, and higher than true FDP for smaller t values, and (iii) the approximation accuracy of the estimated FDP increases with the sample size.

\begin{figure}[!h]
	
	\begin{subfigure}[b]{0.45\textwidth}
		\includegraphics[width=\textwidth, height=0.28\textheight]{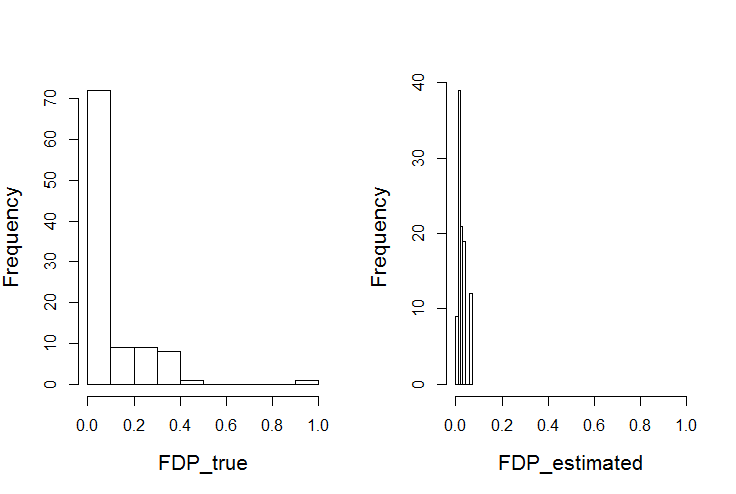}
		\caption{$t = 3.6$ and $n=100$.}
		\label{fig:FDP_f1}
	\end{subfigure}
	\hfill
	\begin{subfigure}[b]{0.45\textwidth}
		\includegraphics[width=\textwidth, height=0.28\textheight]{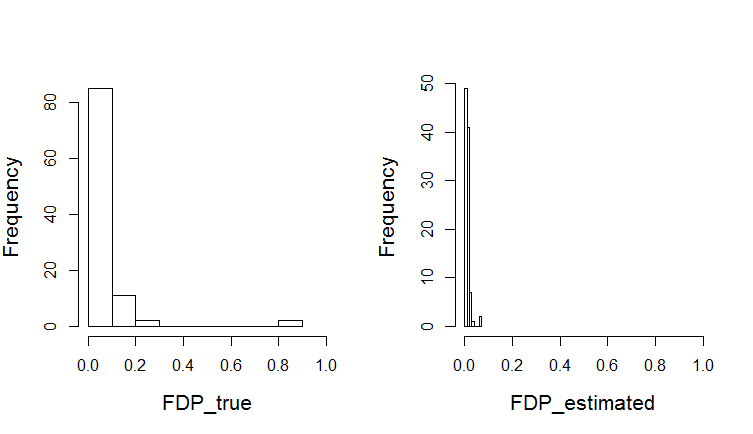}
		\caption{$t = 3.6$ and $n=150$.}
		\label{fig:FDP_f2}
	\end{subfigure}
	\begin{subfigure}[b]{0.45\textwidth}
		\includegraphics[width=\textwidth, height=0.28\textheight]{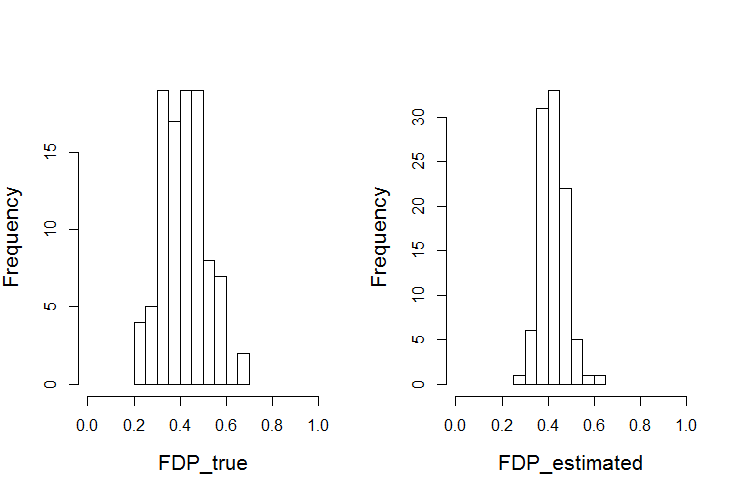}
		\caption{$t = 2$ and $n=100$.}
		\label{fig:FDP_f3}
	\end{subfigure}
	\hfill
	\begin{subfigure}[b]{0.45\textwidth}
		\includegraphics[width=\textwidth, height=0.28\textheight]{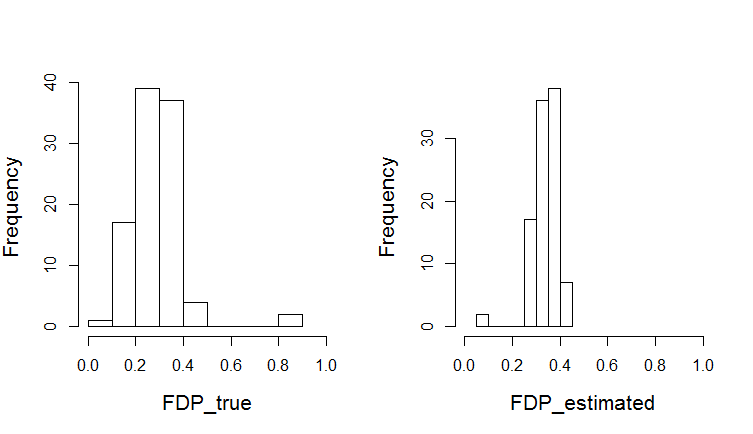}
		\caption{$t = 2$ and $n=150$.}
		\label{fig:FDP_f4}
	\end{subfigure}
\caption{Histograms of the true FDP (\textsf{FDP\_true}) and estimated FDP (\textsf{FDP\_estimated}) when  $p=200$, $\beta_1=0.5$, and $s_0=30$. } \label{fig:hist}
\end{figure}

We also show the histograms of the true FDP and estimated FDP at specific $t$ values with $p=200$, $\beta_1 = 0.5$, $s_0=30$, $n = 100$ or 150. Figure \ref{fig:hist} shows that the distribution of the estimated FDP generally mimics that of the true FDP in a more concentrated way. When sample size increases, the true FDP and the estimated FDP become more concentrated around their own mean values.

\noindent {\bf Example 3: Variable selection by DLasso-FDP procedure}.
We compare DLasso-FDP with three other methods,  DLasso-FWER, DLasso-BH, and Knockoff.
DLasso-FWER is the dependence adjusted FWER control method in \cite{Buhlmann2013} and \cite{vandegeer2014}.
DLasso-BH is an ad hoc procedure that directly applies Benjamini-Hochber's procedure \citep{Benjamini:1995} on the asymptotic $p$-values of the DLasso estimator.
%To our knowledge, the validity of DLasso-BH has not been justified.
The first three methods (DLasso-FDP, DLasso-FWER, and DLasso-BH) are all built upon the DLasso estimator. The fourth method, Knockoff, has been developed to directly control FDR without the need to derive limiting distribution and p-values \citep{Barber15, CandesPanning}. We use the "knockoff.filter" function in default from the \textsf{R} package \texttt{knockoff}, which creates model-X second-order Gaussian knockoffs as introduced in \cite{CandesPanning}.
The nominal levels are set at 0.1 for all the methods.

%Specifically, $s_0=10$ in Figure \ref{fig:rank_4} and  $s_0=40$ in Figure \ref{fig:rank_2}.
The performances of the methods are measured by the mean values of their true FDPs and TPPs from 100 simulations.
Note that the expected value of FDP is FDR.
\autoref{tab:FDP_control_1} has $s_0=10$, $n=100$ and 150, $\beta_1 = 0.5, 0.7,$ and 1.  \autoref{tab:FDP_control_2} has an increased value for $s_0$ to 30.
Both tables show that  DLasso-BH seems to control the empirical FDR the worst and  DLasso-FWER, on the contrary, is most conservative with smallest empirical FDR. For DLasso-FDP, we see that when sample size increases, DLasso-FDP has a better control on the empirical FDR at the nominal level of 0.1, which agrees with our expectation.
Comparing DLasso-FDP with Knockoff, it shows that neither of the two methods dominates the other in all the settings. When $s_0$ is relatively small in \autoref{tab:FDP_control_1}, DLasso-FDP tends to have higher  TPP than Knockoff, especially when coefficient values are small. On the other hand, when $s_0$ is relatively large in \autoref{tab:FDP_control_2} (so that the sparsity condition on $s_0$ in assumption A3) may not hold), Knockoff tends to have higher TPP than DLasso-FDP, especially when coefficient values are relatively large.

\begin{table}[!h]
	\small
	\centering
	
	\begin{tabular}{ccc| c c  c c}
		\hline
		 $n$ & $\beta_1$ & & DLasso-FDP  & DLasso-BH & DLasso-FWER & Knockoff \\
		\hline
		100 & 0.5 & FDP & 0.171 & 0.248 & 0.080 & 0.097 \\
	        &   	& TPP & 0.856 & 0.884 & 0.774 & 0.383 \\
		\cline{4-7}
			& 0.7 & FDP & 0.146 & 0.237 & 0.080 & 0.151 \\
			& 	  & TPP & 0.962 & 0.972 & 0.94 & 0.749 \\
		\cline{4-7}	
			& 1 & FDP &  0.151 & 0.236 & 0.065 & 0.109 \\
		&       & TPP &  0.998  & 0.998 & 0.997 & 0.889 \\
		\hline
	  150   & 0.5 & FDP & 0.090 & 0.152 & 0.037 & 0.111 \\
		    &       & TPP & 0.832 & 0.863 & 0.756 & 0.517 \\
		\cline{4-7}
        	& 0.7 & FDP & 0.064 & 0.104 & 0.018 & 0.102 \\
        	& 	  & TPP & 0.987 & 0.991 & 0.983 & 0.923 \\
        \cline{4-7}
           	& 1   & FDP & 0.084 & 0.134 & 0.048 & 0.099 \\
             &    	& TPP & 0.983 & 0.986 & 0.967 & 0.930 \\
        \hline
	\end{tabular}
\caption{The mean values of FDP and TPP for different variable selection methods with $s_0=10$ and $p=200$.} \label{tab:FDP_control_1}
\end{table}

\begin{table}[!h]
	\small
	\centering
	
\begin{tabular}{ccc| c c  c c}
	\hline
	$n$ & $\beta_1$ & & DLasso-FDP  & DLasso-BH & DLasso-FWER & Knockoff \\
	\hline
	100 & 0.5 & FDP & 0.164 & 0.182 & 0.107 & 0.072 \\
	    &     & TPP & 0.180 & 0.212 & 0.113 & 0.146  \\
	\cline{4-7}
	& 0.7 & FDP & 0.160 & 0.185 & 0.107 & 0.111 \\
	& 	  & TPP & 0.209 & 0.248 & 0.137 & 0.274 \\
	\cline{4-7}	
	& 1 & FDP  & 0.147 & 0.182 & 0.104 & 0.116 \\
    &  	& TPP & 0.229 & 0.271 & 0.153 & 0.372 \\
	\hline	
	150   & 0.5 & FDP & 0.084 & 0.122 & 0.044 & 0.093 \\
	      &    	& TPP & 0.368 & 0.452 & 0.253 & 0.578 \\
	\cline{4-7}
	& 0.7 & FDP & 0.096 & 0.139 & 0.070 & 0.120 \\
	& 	  & TPP & 0.314 & 0.401 & 0.214 & 0.681 \\
	\cline{4-7}	
	& 1   & FDP &  0.052 & 0.106 & 0.026 & 0.117 \\
	&    & TPP & 0.477  & 0.583 & 0.364 & 0.958\\
	\hline
\end{tabular}
\caption{The mean values of FDP and TPP for different variable selection methods with $s=30$ and $p=200$.} \label{tab:FDP_control_2}
\end{table}

\section{Discussion\label{sec:disc}}

%We study the ranking efficiency of DLasso and propose the DLasso-FDP procedure which is built upon the ranking generated by DLasso and controls FDP at a user-specified level.
Theoretical analyses in the paper have focused on  Gaussian random design.
We show that our procedure can consistently estimate the FDP of variable selection as long as the DLasso estimator is asymptotically normal.
Extensions to random design with sub-Gaussian rows or bounded rows can be developed with minor modifications.

We present the optimality of the standardized DLasso in ranking efficiency when the number of true predictors is relatively small, i.e., $s_0 = o(n/\ln p)$.
When the true predictors are relatively dense, i.e., $s_0 \gg n/\ln p$, relevant predictors always intertwine with noise variables on the Lasso solution path even if all predictors are independent (i.e., $\ms{\Sigma} = \mb{I}$), no matter how large $\beta_{\min}$ is \citep{Wainwright:2009, Su17}.
In this case, we expect improved ranking performance based on $\left\{|z_{j}|\right\}_{j=1}^{p}$ because  DLasso mitigates the bias induced by Lasso shrinkage. Numerical analysis in the paper supports the expectation.
Theoretical analyses in the setting with $s_0 \gg n/\ln p$ are scarce but relevant to real applications with dense causal factors. We hope to investigate more in this direction in future research.

Finally, we point out that the computational burden of DLasso-FDP is mainly caused by precision matrix estimation when dimension of the design matrix is large. Using nodewise regression by Lasso, one essentially solves $p$ Lasso problems with sample size n and dimensionality $p-1$. When $p$ is of thousands or more, computation resources for parallel computing would be needed to facilitate the estimation of precision matrix.
Accelerating the computation for precision matrix estimation without loss of accuracy is of great interest for future research.

\section*{Acknowledgments}
Dr. Jeng was partially supported by the NSF Grant DMS-1811360. We thank the Editor, Associate Editor and referees for their helpful comments.
% in the template, appendix, reference all with \section*{}

\section*{Appendix}

In these appendices, we present some lemmas that are needed for the proofs of the results presented in the main paper. Recall $ \sqrt{n} (  \hat{\mb{b}}-\ms{\beta} )  =\mb{w} - \ms{\delta} $, where
$\mb{w} \sim \mathcal{N}_{p} (  0,\sigma^{2}\hat{\ms{\Omega}} )$
conditional on $\mb{X}$. We call $\mb{w}$ the pivotal statistic.
In all the proofs, the arguments are conditional on $\mb{X}$ unless otherwise noted.
The $O_{\Pr}$ or $o_{\Pr}$ bounds for expectations, covariances or cumulative distribution functions are
induced by the random matrix $\hat{\ms{\Omega}}$ as the covariance matrix of $\mb{w}$.

\subsection*{Extra lemmas\label{secExtraLemmas}}

\begin{lemma}\label{lmSmallop}
  Assume A2) and $s_{\max}=o\left(  n/\ln p\right)  $. Then $ \Vert \boldsymbol{\hat{\Omega}}-\boldsymbol{\Sigma
}^{-1} \Vert _{\infty}=o_{P}\left(  1\right)  $. If further A1b) holds, then $\Vert \hat{\ms{\Theta}}\hat{\ms{\Sigma}}-\mb{I}\Vert_{\infty }
=\bop(\lambda _{1})$, both $\min_{1 \le j \le p}\hat{\ms{\Omega}}_{jj}$ and $\max_{1 \le j \le p}\hat{\ms{\Omega}}_{jj}$ are uniformly bounded (in $p$) away from $0$ and $\infty$ with probability tending to $1$,
and $\left\Vert \boldsymbol{\delta}^{\prime}\right\Vert
_{\infty}\le (\sigma \sqrt{C_{\min}})^{-1} \left\Vert \boldsymbol{\delta}\right\Vert _{\infty} $ with probability tending to $1$.
\end{lemma}
\begin{proof}
With A2) and $s_{\max}=o\left(  n/\ln p\right)$, the conditions of Lemmas 5.3 and 5.4 of \cite{vandegeer2014} are satisfied, i.e., $\lambda_{j}$ is of order $\sqrt{n^{-1}\ln p}$ for each $j=1,\ldots,p$, $\max_{1 \le j \le p}s_j=o\left(  n/\ln p\right)  $ and $\max_{1 \le j \le p}\lambda_{j}^{2}s_{j}=o\left(1\right)  $. So, $ \Vert \boldsymbol{\hat{\Omega}}-\boldsymbol{\Sigma}^{-1} \Vert _{\infty}=o_{P}\left(  1\right)  $.

Note that for the positive definite matrix $\ms{\Omega} = \ms{\Sigma}^{-1}$, the largest and smallest among $\ms{\Omega}_{jj}$ for $j=1,\ldots,p$ are sandwiched between $C_{\min}$ and $C_{\max}$.
If in addition A1b) holds, then
$\hat{\ms{\Omega}}_{jj}, j=1,\ldots,p$ are uniformly bounded away from $0$ and
$\infty$ with probability tending to $1$, inequality (10) of \cite{vandegeer2014} implies $\Vert \hat{\ms{\Theta}}\hat{\ms{\Sigma}}-\mb{I}\Vert_{\infty }=\bop(\lambda _{1})$, and $\left\Vert \boldsymbol{\delta}^{\prime}\right\Vert
_{\infty}\le (\sigma \sqrt{C_{\min}})^{-1} \left\Vert \boldsymbol{\delta}\right\Vert _{\infty} $ with probability tending to $1$. This completes the proof.
\end{proof}

%% Keep this remark in order to check the above paragraph
%\noindent {\bf Remark}.   Lemma 5.3 of \cite{vandegeer2014} proved that $ \Vert \boldsymbol{\hat{\Omega}}-\boldsymbol{\Sigma
%}^{-1} \Vert _{\infty}=o_{P}\left(  1\right)  $ under the assumptions A1),
%$\lambda_{j}\asymp\sqrt{n^{-1}\ln p}$, $\max_{j}\lambda_{j}^{2}s_{j}=o\left(
%1\right)  $ and $\max_j s_{j}=o\left(  n/\ln p\right)  $. Under the conditions here, these assumptions
%are satisfied, meaning that $ \Vert \boldsymbol{\hat{\Omega}}-\boldsymbol{\Sigma
%}^{-1} \Vert _{\infty}=o_{P}\left(  1\right)  $ is still valid.

\begin{lemma}\label{LmOmegaL1}
  Let $\hat{\mb{K}}$ be the correlation matrix of $\mb{w}$.  Assume A1) and A2). Then
  \begin{equation}\label{eqCorMatL1}
    p^{-2}\Vert \sigma^{2}\hat{\ms{\Omega}}\Vert _{1} = \bop\left(  \lambda_{1}\sqrt{s_{\max}}\right)
    \quad \text{and} \quad \Vert \hat{\mb{K}}\Vert _{1} = O( \sigma^{2}\Vert \hat{\ms{\Omega}}\Vert _{1}).
  \end{equation}
\end{lemma}
\begin{proof}
Recall $\hat{\ms{\Omega}}=\hat{\ms{\Theta}}\hat{\ms{\Sigma}}\hat{\ms{\Theta}}^\top$, the covariance matrix of $\mb{w}$.
Since $\sigma$ is
bounded, then $\Vert\sigma^{2} \hat{\ms{\Omega}}\Vert _{1}={O} (\Vert \hat{\ms{\Omega}}\Vert _{1} )  $.
Recall $\hat{\ms{\theta}}_{j}$ is the $j$th row of $\hat{\ms{\Theta}}$. By triangular
inequality,
\begin{equation}
\Vert \hat{\ms{\Omega}}\Vert_{1} \leq \Vert (\hat{\ms{\Theta}}\hat{%
\ms{\Sigma}}-\mb{I})\hat{\ms{\Theta}}^\top\Vert_{1}+\Vert \hat{\ms{\Theta}}%
^\top\Vert_{1}\leq \sum_{j=1}^{p}\Vert (\hat{\ms{\Theta}}\hat{\ms{\Sigma}}%
-\mb{I})\hat{\ms{\theta}}_{j}^\top\Vert _{1}+\sum_{j=1}^{p}\Vert \hat{\ms{\theta}}%
_{j}\Vert_{1}.  \label{0}
\end{equation}%
To bound $\Vert \hat{\ms{\Omega}}\Vert_{1}$, we bound $\Vert
\hat{\ms{\theta}}_{j}\Vert_{1}$ and $\Vert (\hat{\ms{\Theta}}\hat{\ms{\Sigma}}%
-\mb{I})\hat{\ms{\theta}}_{j}^\top\Vert_{1}$ separately.
First,
\begin{equation*}
\Vert \hat{\ms{\theta}}_{j}\Vert_{1} \le  \Vert \hat{\ms{\theta}}_{j} - \ms{\theta}_{j} \Vert_{1} + \Vert \ms{\theta}_{j}\Vert_{1}
\label{theta_j}.
\end{equation*}%
By Theorem 2.4 of \cite{vandegeer2014}, $\Vert \hat{\ms{\theta}}_{j} - \ms{\theta}_{j} \Vert_{1}  = O_{\Pr}(s_j \lambda_j)$. By Cauchy-Schwartz inequality,  $\Vert \ms{\theta}_{j}\Vert_{1} \le \sqrt{s_{j}} \Vert \ms{\theta}_{j}\Vert_{2} $, and from the discussion in paragraph 5 on page 1178 of \cite{vandegeer2014}, we see $\Vert \ms{\theta}_{j}\Vert_{2}\le C_{\min}^{-2} = O(1) $. Since $s_j \lambda_j \ll \sqrt{s_j}$, then
\begin{equation}
\Vert \hat{\ms{\theta}}_{j}\Vert_{1} \le {O}_{P}(\sqrt{s_{j}})
\label{theta_j}.
\end{equation}%
Next consider $\Vert (\hat{\ms{\Theta}}\hat{\ms{\Sigma}}-\mb{I})\hat{\ms{\theta}}%
_{j}^\top\Vert_{1}$ for any $j=1,\ldots p$. By \autoref{lmSmallop}, we have
$\Vert \hat{\ms{\Theta}}\hat{\ms{\Sigma}}-\mb{I}\Vert_{\infty }
=\bop(\lambda _{1})$.
This, together with (\ref{theta_j}), gives
\begin{equation}
\Vert (\hat{\ms{\Theta}}\hat{\ms{\Sigma}}-\mb{I})\hat{\ms{\theta}}_{j}^\top\Vert
_{1}\leq p\Vert \hat{\ms{\Theta}}\hat{\ms{\Sigma}}-\mb{I}\Vert_{\infty
}\Vert \hat{\ms{\theta}}_{j}\Vert_{1} = \bop(p\lambda
_{j})\bop(\sqrt{s_{j}})=\bop(p\lambda _{j}\sqrt{s_{j}}).  \label{0.1}
\end{equation}
Combing (\ref{theta_j}) and (\ref{0.1}) with (\ref{0}) gives
\begin{equation*}
\Vert \hat{\ms{\Omega}}\Vert_{1} = \bop(p^{2}\lambda _{j}\sqrt{%
s_{j}})+\bop(p\sqrt{s_{j}})= \bop(p^{2}\lambda _{j}\sqrt{s_{j}}).
\end{equation*}
Since $\lambda_j$'s are of the same order by assumption A2), we have $p^{-2}\Vert \sigma^{2}\hat{\ms{\Omega}
}\Vert _{1} = \bop\left(  \lambda_{1}\sqrt{s_{\max}}\right)  $, which is the fist part of
\eqref{eqCorMatL1}.

By \autoref{lmSmallop}, $\Vert \sigma^{2}\hat{\ms{\Omega}}\Vert_{1}=O (\Vert \hat{\mb{K}}\Vert _{1} )$ and
the second part of \eqref{eqCorMatL1} holds. This completes the proof.
\end{proof}

\begin{lemma}\label{LmDiffExpectations}
	Assume A1) to A3). Then
	\begin{equation}\label{eqNexp}
	\vert \E\{\bar{V}_{\mb{z}}\left(  t\right)\}-\E\{\bar{V}_{\tilde{\mb{w}}}\left(  t\right)\} \vert = \sop(1)
	\quad \text{and} \quad
	\vert \bar{V}_{\mb{z}}\left(  t\right)  -\bar{V}_{\tilde{\mb{w}}}\left(t\right) \vert = \sop(1).
	\end{equation}
	Further, $\Var\{\bar V_{\mb{z}} \left(t\right) \} - \Var\{\bar V_{\tilde{\mb{w}}} \left(t\right) \} = \sop(1)$.
\end{lemma}
\begin{proof}

	For $i \in I_{0}$, let $F_{p,i}$ be CDF of $z_{i}$ and $\Phi_{p,i}$ that of
	$w_{i}^{\prime}$. Note that $\beta_i=0$ for all $i \in I_{0}$ and that each $w_{i}^{\prime}$ has unit variance conditional on $\hat{\ms{\Omega}}$. Recall $\ms{\Theta} = \ms{\Sigma}^{-1}$.
	By \autoref{lmSmallop}, $\Vert \hat{\ms{\Omega}}-\ms{\Theta}\Vert _{\infty}=\sop \left(	1\right)  $.
	So, with probability approaching to $1$, $\tilde{\mb{w}}$ has a nondegenerate multivariate Normal (MVN) distribution, and
	$\Phi_{p,i}$ is absolutely continuous with respect to the Lebesgue measure on $\mbr$ for any $1\leq i<j\leq p$.
	Further, $\Vert \ms{\delta}^{\prime} \Vert_{\infty}=\sop(1)$ in view of \autoref{lmJM16} and \autoref{lmSmallop}. Therefore, for any $x \in \mbr$,
	\begin{equation}\label{eqDiffcdf}
	\max_{i \in I_{0}}\left\vert F_{p,i}\left(  x\right)  -\Phi_{p,i}\left(  x\right)  \right\vert = \sop(1).
	\end{equation}
	Let $F_{p,i,j}$ be the joint CDF of $(z_i,z_j)$ and $\Phi_{p,i,j}$ that of $(w^{\prime}_i,w_j^{\prime})$ for each distinct pair of $i$ and $j$.
	Then, for any $x,y \in \mathbb{R}$, we have
	\begin{equation}\label{eq2Dintegral}
	\max_{i \ne j; i,j \in I_{0}}\left\vert F_{p,i,j}(x,y) - \Phi_{p,i,j}(x,y) \right\vert = \sop(1).
	\end{equation}
	Therefore, by \eqref{eqDiffcdf}, the first equality in \eqref{eqNexp} holds.
	Let $$\zeta_{p}(t) = \max_{i\in I_{0}}\left\vert 1_{\left\{  \left\vert z_{i}%
		\right\vert \leq t\right\}  }-1_{\left\{  \left\vert w^{\prime}_{i}\right\vert \leq
		t\right\}  }\right\vert .$$
	Then \eqref{eqDiffcdf} implies $\zeta_{p}(t)=\sop(1)$, and the second equality in \eqref{eqNexp} holds.
	
	Now we show the last claim. Clearly,
	\[
	\Var\{\bar V_{\mb{z}} \left(t\right) \} = \frac{1}{p_0^2} \sum_{j \in I_0} \Var(1_{\{|w_j^{\prime} - \delta_j^{\prime}|>t\}})  + \frac{1}{p_0^2} \sum_{i \ne j; i,j \in I_{0}} \Cov(1_{\{|w_i^{\prime} - \delta_i^{\prime}|>t\}}, 1_{\{|w_j^{\prime} - \delta_j^{\prime}|>t\}})
	\]
	and the first summand in the above identity is $o(1)$ when $p_0 \to \infty$.
	However, \eqref{eqDiffcdf} and \eqref{eq2Dintegral} imply that
	\begin{equation*}%\label{eqDiffCov}
	\max_{i \ne j; i,j \in I_{0}}\left\vert \Cov (1_{\{|w_i^{\prime} - \delta_i^{\prime}|>t\}}, 1_{\{|w_j^{\prime} - \delta_j^{\prime}|>t\}}) - \Cov (1_{\{|w_i^{\prime}|>t\}}, 1_{\{|w_j^{\prime}|>t\}}) \right\vert= \sop(1).
	\end{equation*}
	Thus, $\Var\{\bar V_{\mb{z}} \left(t\right) \} - \Var\{\bar V_{\tilde{\mb{w}}} \left(t\right) \} = \sop(1)$.
	This completes the proof.
\end{proof}

\subsection*{Proof of \autoref{ThmSeparation}}

Recall $\sqrt{n} (  \mathbf{\hat{b}}-\boldsymbol{\beta} )
=\mathbf{w} - \boldsymbol{\delta}$, where $\mathbf{w}|\mathbf{X}\sim
\mathcal{N}_{p} (  0,\sigma^{2}\boldsymbol{\hat{\Omega}} )$.
Let $\mu_{j}=\frac{\sqrt{n}\beta_{j}%
}{\sigma\sqrt{\boldsymbol{\hat{\Omega}}_{jj}}}$, $w_{j}^{\prime}=\frac{w_{j}%
}{\sigma\sqrt{\boldsymbol{\hat{\Omega}}_{jj}}}$ and $\delta_{j}^{\prime}%
=\frac{\delta_{j}}{\sigma\sqrt{\boldsymbol{\hat{\Omega}}_{jj}}}$ for each $j$.
Then%
\begin{equation} \label{eq:z_j}
z_{j}=\mu_{j}+w_{j}^{\prime}-\delta_{j}^{\prime}
\end{equation}
and each $w_{j}^{\prime}$ has unit variance. Set $\tilde{\mb{w}} = \left( w_{1}^{\prime}%
,\ldots,w_{p}^{\prime}\right)  ^\top$ and $\boldsymbol{\delta}^{\prime}=\left(  \delta_{1}^{\prime}%
,\ldots,\delta_{p}^{\prime}\right)  ^\top$.

By \autoref{lmSmallop}, $\left\Vert \boldsymbol{\delta}^{\prime}\right\Vert
_{\infty}\le (\sigma \sqrt{C_{\min}})^{-1} \left\Vert \boldsymbol{\delta}\right\Vert _{\infty} $ with probability tending to $1$.
So, \autoref{lmJM16} implies
\begin{equation}\label{eq12a}
\Pr\left\{\left\Vert \boldsymbol{\delta}^{\prime}\right\Vert _{\infty} >
(\sigma \sqrt{C_{\min}})^{-1} B_{p}\left(  s_{0},n,\boldsymbol{\Sigma}\right)\right\} \to 0,
\end{equation}
where we recall
$$B_{p}\left(  s_{0},n,\boldsymbol{\Sigma}\right)  =c^{\prime}\rho\sigma
\sqrt{\frac{s_{0}}{n}}\ln p+c^{\prime}\sigma\min\left\{  s_{0},s_{\max}\right\}
\frac{\ln p}{\sqrt{n}}.$$
For simplicity, we will denote $B_{p}\left(  s_{0},n,\boldsymbol{\Sigma}\right)$ by $B_p$.

Now we break the rest of the proof into two steps: bounding $\max_{j\in I_{0}}\left\vert
w_{j}^{\prime} - \delta_j^{\prime}\right\vert $ from above and bounding $\min_{i\in S_{0}}\left\vert \mu
_{j}+w_{j}^{\prime} - \delta_j^{\prime}\right\vert $ from below.

{\bf Step 1:} bounding $\max_{j\in I_{0}}\left\vert
w_{j}^{\prime} - \delta_j^{\prime}\right\vert $ from above. Recall $C_{p}=\ln (p^{2} / 2\pi) +\ln\ln (p^{2} / 2\pi)$ and let%
\begin{equation*}
Q_{p}=C_{p}+2\mathcal{G},%\label{eq14}%
\end{equation*}
where $\mathcal{G}$ is an exponential random variable with expectation $1$. From Theorem 3.3 of
\cite{hartigan2014}, we obtain%
\begin{equation*}
\max_{j\in I_{0}}\left\vert w_{j}^{\prime}\right\vert
^{2}\leq Q_{p_{0}}%\label{eq8}%
\end{equation*}
with probability tending to $1$ as $p_0\rightarrow\infty$. This, together with \eqref{eq12a}, implies%
\begin{equation*} %\label{eq6}
\max_{j\in I_{0}}\left\vert
w_{j}^{\prime} - \delta_j^{\prime}\right\vert  \le  \sqrt{Q_{p_{0}}} + (\sigma \sqrt{C_{\min}})^{-1} B_p
\end{equation*}
with probability tending to $1$ as $p_0 \to \infty$.

{\bf Step 2:} bounding $\min_{i\in S_{0}}\left\vert \mu_{j}+w_{j}^{\prime
} - \delta_j^{\prime}\right\vert $ from below. Applying Theorem 3.3 of \cite{hartigan2014} to
$\max_{j\in S_{0}}\left\vert w_{j}^{\prime}\right\vert $ and noticing
$s_{0}\leq p_{0}$, we obtain%
\begin{equation}  \label{2.1}%
\max_{j\in S_{0}}\left\vert w_{j}^{\prime}\right\vert \leq\sqrt{Q_{s_{0}}}%
\leq\sqrt{Q_{p_{0}}}%.
\end{equation}
with probability tending to $1$ as $s_0 \to \infty$.
So, (\ref{eq12a}) and (\ref{2.1}) imply
\begin{equation*}  %\label{eq22x}%
\min_{j\in S_{0}}\left\vert \mu_{j}+w_{j}^{\prime} - \delta_j^{\prime}\right\vert
%\geq
%\min_{j\in S_{0}}\left\vert \mu_{j} \right\vert
%- \max_{j\in S_{0}} \left\vert w'_{j} \right\vert
%- \max_{j\in S_{0}} \left\vert \delta_j^{\prime}\right\vert
\geq
\min_{j\in S_{0}}\left\vert \mu_{j} \right\vert - \sqrt{Q_{p_{0}}} - (\sigma \sqrt{C_{\min}})^{-1} B_p
\end{equation*}
with probability tending to $1$ as $s_{0}\rightarrow\infty$.

Finally, we show the separation between the relative predictors and irrelevant ones. Consider the probability:
\begin{eqnarray*}
	& & \Pr\left\{\min_{j\in S_{0}}\left\vert \mu_{j} \right\vert - \sqrt{Q_{p_{0}}} - (\sigma \sqrt{C_{\min}})^{-1} B_p \le \sqrt{Q_{p_{0}}} + (\sigma \sqrt{C_{\min}})^{-1} B_p \right\} \\
	& = & \Pr\left\{\sqrt{Q_{p_{0}}} \ge 2^{-1} \min_{j\in S_{0}}\left\vert \mu_{j} \right\vert - (\sigma \sqrt{C_{\min}})^{-1} B_p \right\} \\
	& = & \Pr\left\{\sqrt{C_{p}+2\mathcal{G}} \ge 2^{-1} \min_{j\in S_{0}}\left\vert \mu_{j} \right\vert - (\sigma \sqrt{C_{\min}})^{-1} B_p \right\}.
\end{eqnarray*}
Then, the above probability converges to $0$ as $s_0 \to \infty$ if $$2^{-1} \min_{j\in S_{0}}\left\vert \mu_{j} \right\vert - (\sigma \sqrt{C_{\min}})^{-1} B_p \ge (1+a) \sqrt{C_p}$$ for some constant $a>0$, for which the last inequality holds when
\[
\min_{j\in S_{0}} \left\vert \beta_{j}\right\vert \geq 2  n^{-1/2} \left\{ \sqrt{C_{\min}^{-1}C_{\max}}B_{p}\left(  s_{0},n,\boldsymbol{\Sigma}\right) + \sigma \sqrt{C_{\max}} (1+a)\sqrt{C_{p_{0}}}\right\}.
\]
This completes the proof.

\subsection*{WLLN for multiple testing based on the pivotal statistic}

From \autoref{LmOmegaL1}, we can obtain a ``weak law of large numbers (WLLN)" for
$\{\bar{R}_{\tilde{\mb{w}}}\left(  t\right)\}_{p \ge 1}$ and $\{\bar{V}_{\tilde{\mb{w}}}\left(  t\right)\}_{p \ge 1}$.
To achieve this, we need some facts on Hermite polynomials and Mehler expansion
since they will be critical to proving \autoref{PropOrderCov}.
Let $\phi\left(  x\right)
=\left(  2\pi\right)  ^{-1/2}\exp\left(  -x^{2}/2\right)  $ and
\[
f_{\rho}\left(  x,y\right)  =\frac{1}{2\pi\sqrt{1-\rho^{2}}}\exp\left\{
-\frac{x^{2}+y^{2}-2\rho xy}{2\left(  1-\rho^{2}\right)  }\right\}
\]
for $\rho\in\left(  -1,1\right)$.
For a nonnegative integer $k$, let $H_{k}\left(  x\right)  =\left(  -1\right)  ^{k}\frac{1}{\phi\left(
	x\right)  }\frac{d^{k}}{dx^{k}}\phi\left(  x\right)  $ be the $k$th Hermite
polynomial; see \cite{Feller:1971B} for such a definition. Then
Mehler's expansion \citep{Mehler:1866} gives%
\begin{equation}
f_{\rho}\left(  x,y\right)  =\left\{  1+\sum\nolimits_{k=1}^{\infty}\frac{\rho^{k}}%
{k!}H_{k}\left(  x\right)  H_{k}\left(  y\right)  \right\}  \phi\left(
x\right)  \phi\left(  y\right). \label{eq:Mehler}%
\end{equation}
Further, Lemma 3.1 of \cite{chen2016} asserts%
\begin{equation}
\left\vert e^{-y^{2}/2}H_{k}\left(  y\right)  \right\vert \leq C_{0}\sqrt
{k!}k^{-1/12}e^{-y^{2}/4}\text{ \ for any\ }y\in\mathbb{R}
\label{eqBoundHermite}%
\end{equation}
for some constant $C_{0}>0$.

With the above preparations, we have:
%% intended for WLLN
\begin{lemma}\label{PropOrderCov}
Assume A1) and A2). Then
\begin{equation}\label{eqOrderW}
 \Var \{  \bar{R}_{\tilde{\mb{w}}}\left(  t\right)   \}=\bop\left(\max\{p^{-1},\lambda_{1}\sqrt{s_{\max}}\}\right); \quad
   \Var \{  \bar{V}_{\tilde{\mb{w}}}\left(  t\right)  \}=\bop\left(\max\{p_0^{-1},\lambda_{1}\sqrt{s_{\max}}\}\right).
\end{equation}
If in addition assumption A3) is valid, then
\begin{equation}\label{SLLNNull}
\vert \bar{R}_{\tilde{\mb{w}}}\left(  t\right)-\E \{\bar{R}_{\tilde{\mb{w}}}\left(  t\right)\} \vert = \sop(1)
\quad \text{and} \quad
\vert \bar{V}_{\tilde{\mb{w}}}\left(  t\right)-\E \{\bar{V}_{\tilde{\mb{w}}}\left(  t\right)\} \vert = \sop(1).
\end{equation}
\end{lemma}

\begin{proof}
Let $\rho_{ij}$ be the correlation between $w_{i}^{\prime}$ and $w_{j}^{\prime}$ for $i \neq j$. Define sets%
\begin{equation}
\left\{
\begin{array}
[c]{c}%
B_{1,p}=\left\{  \left(  i,j\right)  :1\leq i,j\leq p,i\neq j,\left\vert
\rho_{ij}\right\vert <1\right\}  ,\\
B_{2,p}=\left\{  \left(  i,j\right)  :1\leq i,j\leq p,i\neq j,\left\vert
\rho_{ij}\right\vert =1\right\}  .\nonumber
\end{array}
\right.  %\label{eqSets}%
\end{equation}
Namely, $B_{2,p}$ is the set of distinct pair $\left(  i,j\right)$ such that $w_{i}^{\prime}$ and $w_{j}^{\prime}$ are linearly dependent.
Let
$C_{\tilde{\mb{w}},ij}=\Cov \left(  1_{\left\{  \vert w_{i}^{\prime}\vert \leq t\right\}
},1_{\left\{  \vert w_{j}^{\prime}\vert \leq t\right\}  }\right)  $ for $i \neq j$. Then%
\begin{equation}\label{eqCovExpand}
\Var \{  \bar{R}_{\tilde{\mb{w}}}\left(  t\right)   \}  =p^{-2}\sum_{j=1}%
^{p}\Var \left(  1_{\left\{  \left\vert w_{j}^{\prime}\right\vert \leq t\right\}
} \right)  +p^{-2}\sum_{(i,j)\in B_{1,p}}C_{\tilde{\mb{w}},ij} + p^{-2}\sum_{(i,j)\in B_{2,p}}C_{\tilde{\mb{w}},ij}.
\end{equation}
Since
\[
 p^{-2}\sum\nolimits_{(i,j)\in B_{2,p}} \vert C_{\tilde{\mb{w}},ij} \vert  = O(p^{-2} \vert B_{2,p} \vert)
 = O (p^{-2} \Vert \hat{\mb{K}} \Vert_1 )
\]
and
\[
p^{-2}\sum_{j=1}^{p} \Var \left(  1_{\left\{  \left\vert w_{j}^{\prime}\right\vert
\leq t\right\}  } \right)  =O (p^{-1} ),
\]
\eqref{eqCovExpand} becomes
\begin{equation} \label{eq:1}
\Var \{  \bar{R}_{\tilde{\mb{w}}}\left(  t\right)   \}  = O (p^{-1} ) +
O (p^{-2} \Vert \hat{\mb{K}} \Vert_1 ) + p^{-2}\sum\nolimits_{(i,j)\in B_{1,p}}C_{\tilde{\mb{w}},ij}.
\end{equation}

Consider the last term on the right hand side of \eqref{eq:1}. Define $c_{1,i}=-t$ and $c_{2,i}=t$.
Fix a pair of $(i,j)$ such that $i \neq j$ and $\vert \rho_{ij} \vert \neq 1$.
Since $C_{\tilde{\mb{w}},ij}$ is finite and the series in Mehler's expansion in (\ref{eq:Mehler})
as a trivariate function of $\left(  x,y,\rho\right)  $ is uniformly
convergent on each compact set of $\mathbb{R}\times\mathbb{R}\times\left(
-1,1\right)  $ as justified by \cite{Watson:1933}, we can interchange the order the summation and integration and obtain%
\begin{eqnarray*}
C_{\tilde{\mb{w}},ij}  & = & \int_{c_{1,i}}^{c_{2,i}} \int_{c_{1,j}}^{c_{2,j}}  f_{\rho_{ij}}\left(  x,y\right) dx dy -  \int_{c_{1,i}}^{c_{2,i}}  \phi(x) dx  \int_{c_{1,j}}^{c_{2,j}} \phi(y) dy \\
& = & \sum_{k=1}^{\infty}\frac{\rho_{ij}^{k}}{k!} \int_{c_{1,i}}^{c_{2,i}}  H_k(x) \phi(x) dx  \int_{c_{1,j}}^{c_{2,j}} H_k(y) \phi(y) dy.
\end{eqnarray*}
Since
$H_{k-1}\left(  x\right)  \phi\left(  x\right)  =\int_{-\infty}^{x}%
H_{k}\left(  y\right)  \phi\left(  y\right)  dy$ for $x\in\mathbb{R}$, then
\[
C_{\tilde{\mb{w}},ij} = \sum_{k=1}^{\infty}\frac{\rho_{ij}^{k}}{k!}
\{H_{k-1}(c_{2,i}) \phi(c_{2,i}) - H_{k-1}(c_{1,i}) \phi(c_{1,i})\}\{H_{k-1}(c_{2,j}) \phi(c_{2,j}) - H_{k-1}(c_{1,j}) \phi(c_{1,j})\}.
\]
Therefore,
\begin{equation*} %\label{eq:2}
\left \vert p^{-2}\sum\nolimits_{(i,j)\in B_{1,p}}C_{\tilde{\mb{w}},ij}\right \vert \le \sum_{l,l^{\prime}\in\left\{  1,2\right\}  }
\Psi_{p,l,l^{\prime}}^{\ast},
\end{equation*}
where%
\[
\Psi_{p,l,l^{\prime}}^{\ast}= p^{-2}\sum_{1\leq i<j\leq p}\sum_{k=1}^{\infty}\frac{\left\vert
	\rho_{ij}\right\vert ^{k}}{k!}\left\vert H_{k-1}\left(  c_{l,i}\right)
\phi\left(  c_{l,i}\right)  H_{k-1}\left(  c_{l^{\prime},j}\right)  \phi\left(
c_{l^{\prime},j}\right)  \right\vert
\]
for $l,l^{\prime}\in\left\{  1,2\right\}  $. For any fixed pair $\left(
l,l^{\prime}\right)  $, inequality (\ref{eqBoundHermite}) implies%
\begin{equation*}
\Psi_{p,l,l^{\prime}}^{\ast}  \le p^{-2}\sum_{1\leq i<j\leq p}\left\vert
\rho_{ij}\right\vert \sum_{k=1}^{\infty}k^{-7/6}\left\vert \rho_{ij}%
\right\vert ^{k-1}
\exp\left(  -c_{l,i}^{2}/4\right)  \exp\left(  -c_{l^{\prime},j}%
^{2}/4\right).% \label{eqVar2}\\
\end{equation*}
So,
\begin{equation}\label{eqSbnd4}
\Psi_{p,l,l^{\prime}}^{\ast}\leq p^{-2}\sum_{1\leq i<j\leq p}\left\vert
\rho_{ij}\right\vert ={O} ( p^{-2}\Vert \hat{\mb{K}}\Vert _{1} ),
\end{equation}
which, together with (\ref{eqSbnd4}), implies
\begin{equation} \label{eq:3}
\left \vert p^{-2}\sum\nolimits_{(i,j)\in B_{1,p}}C_{\tilde{\mb{w}},ij} \right \vert = O\left( p^{-2}\Vert \hat{\mb{K}}\Vert _{1}\right).
\end{equation}
Combing (\ref{eq:1}) and (\ref{eq:3}) with the result $\Vert p^{-2}\hat{\mb{K}}\Vert _{1}=\bop\left(\lambda_1 \sqrt{s_{\max}}\right)$ from \autoref{LmOmegaL1} gives
\begin{equation}\label{eqSbnd5}
\Var \{  \bar{R}_{\tilde{\mb{w}}}\left(  t\right)   \} = O\left(p^{-1}\right) + \bop\left(\lambda_1 \sqrt{s_{\max}}\right).
\end{equation}
By restricting the expansion on the right hand side of \eqref{eqCovExpand} to the index set $(i,j) \in I_{0} \times I_{0}$ for $i \neq j$ and to $I_{0}$ for $j$, changing $p$ there into $p_0$, and following almost identical arguments that lead to \eqref{eqSbnd5}, we see that
$\Var \{  \bar{V}_{\tilde{\mb{w}}}\left(  t\right)   \} = O\left(p_0^{-1}\right) + \bop\left(\lambda_1 \sqrt{s_{\max}}\right)$.
Therefore, \eqref{eqOrderW} holds.
Finally, applying Chebyshev inequality to $\bar{R}_{\tilde{\mb{w}}}\left(
t\right)  $ and $\bar{V}_{\tilde{\mb{w}}}\left(  t\right)  $ with the bounds in \eqref{eqOrderW} gives \eqref{SLLNNull}. This completes the proof.
\end{proof}

\subsection*{Proof of \autoref{ThmSLLNFDP}}

Recall the decomposition of $z_j$ in (\ref{eq:z_j}),  $R_{\mb{z}}\left(  t\right)  =\sum_{j=1}^{p}1_{\left\{  \left\vert z_{j}\right\vert >t\right\}  }$, and $V_{\mb{z}}\left(  t\right) =\sum_{j\in I_{0}} 1_{\left\{  \left\vert z_{j}\right\vert >t\right\}  }$.
Define
$R_{\tilde{\mb{w}}}\left(  t\right)  =\sum_{j=1}^{p}1_{\left\{  \left\vert w_{j}^{\prime}\right\vert >t\right\}  }$
and
$V_{\tilde{\mb{w}}}\left(  t\right)  =\sum_{j\in I_{0}}1_{\left\{  \left\vert w_{j}^{\prime}\right\vert >t\right\}  }$.
Further, define the following averages:
\[
\bar{R}_{\mb{z}}\left(  t\right) = p^{-1 }R_{\mb{z}}\left(  t\right);
\qquad \bar{R}_{\tilde{\mb{w}}}\left(
t\right)  =p^{-1}R_{\tilde{\mb{w}}}\left(  t\right); \qquad \bar{V}_{\mb{z}}\left(
t\right)  =p_{0}^{-1}V_{\mb{z}}\left(  t\right);
\qquad \bar{V}_{\tilde{\mb{w}}}\left(
t\right)  =p_{0}^{-1}V_{\tilde{\mb{w}}}\left(  t\right).
\]
From \autoref{LmDiffExpectations} and \autoref{PropOrderCov}, we have $\left\vert \bar{V}_{\mb{z}}\left(  t\right)- \bar{V}_{\tilde{\mb{w}}}\left(  t\right)  \right\vert =\sop\left(  1\right)$ and $\left\vert \bar{V}_{\tilde{\mb{w}}}\left(  t\right)  -\E\left\{  \bar{V}_{\tilde{\mb{w}}}\left(  t\right)  \right\}\right\vert =\sop\left(  1\right)$. So,
\begin{equation}\label{eq:4}
\left\vert \bar{V}_{\mb{z}}\left(  t\right)  -\E\left\{  \bar{V}_{\tilde{\mb{w}}}\left(  t\right)  \right\}\right\vert =\sop\left(  1\right).
\end{equation}

Next, we show that $\bar{R}_{\mb{z}}(t)$ is bounded away from $0$ uniformly in $p$ with probability tending to $1$.
By their definitions,  $\bar{R}_{\mb{z}}(t) \geq\left(  p^{-1}p_{0}\right) \bar{V}_{\mathbf{z}}\left(  t\right) $ almost surely,  and $p^{-1}p_{0}$ is uniformly bounded in $p$ from below by a positive constant
$\pi_{\ast}$. Then
\[
\Pr\left[ \bar{R}_{\mathbf{z}}\left(  t\right)  > 2^{-1}\pi_{\ast}\E\left\{  \bar{V}_{\tilde{\mb{w}}}\left(  t\right)  \right\} \right]  \rightarrow1,
\]
where
$
\E\left\{  \bar{V}_{\tilde{\mb{w}}}\left(  t\right)  \right\} = 2p_{0}^{-1}%
\sum\nolimits_{j\in I_{0}}\Phi\left(  -t\right) = 2 \Phi\left(  -t\right).
$
Therefore,
\begin{equation}
\Pr\left\{ \bar{R}_{\mathbf{z}}\left(  t\right)  >\pi_{\ast}\Phi\left(
-t\right)  \right\}  \rightarrow1.\label{eq23b}%
\end{equation}

Combining (\ref{eq:4}) and (\ref{eq23b}) gives
\[
\left\vert \frac{V_{\mb{z}}\left(  t\right)  }{R_{\mb{z}}\left(  t\right)  }%
-\frac{\E\left\{V_{\tilde{\mb{w}}}\left(  t\right)  \right\}}{R_{\mb{z}}\left(  t\right)  }\right\vert
=o_{P}\left(  1\right),
\]
and the result in (\ref{eqSLLNfdpA}) follows since $p-p_0=s_0$ and $s_0 / p = o(1)$.
This completes the proof.

\subsection*{Proof of \autoref{Thm:mFDR}}

By \eqref{eq23b}, $\bar{R}_{\mb{z}}\left(  t\right)$ is bounded away from $0$ uniformly in $p$ with probability tending to $1$. So, it suffices to show
\begin{equation}\label{eq4}
\frac{\E \{ V_{\tilde{\mb{w}}}\left(  t\right)   \}  }{R_{\mb{z}}\left(t\right)}  - \frac{\E\{V_{\mb{z}}\left(  t\right)\}  }{	\E\{R_{\mb{z}}\left(  t\right)\}  }
 = \sop(1).
\end{equation}
Since $\E\{\bar V_{\mb{z}}\left(  t\right)\} - \E \{ \bar V_{\tilde{\mb{w}}}\left(  t\right)   \} = \sop(1)$ from \autoref{LmDiffExpectations}, \eqref{eq4} follows once we show
\begin{equation} \label{2}
\bar R_{\mb{z}}\left(t\right) - \E\{\bar R_{\mb{z}}\left(  t\right) \}= \sop(1).
\end{equation}
To this end, we only need to show $\Var\{\bar R_{\mb{z}}\left(t\right)\} = \sop(1)$, which implies \eqref{2}.

%Recall $S_{0}=\left\{j:\beta_{j}\neq0\right\}$ and $s_0 = \left\vert S_{0}\right\vert = p - p_0$.
Observe
\begin{equation}\label{eq5}
\bar R_{\mb{z}}\left(t\right) =
\frac{p_0}{p} \bar{V}_{\mb{z}}(t)
+ \frac{s_0}{p} \frac{1}{s_0} \sum_{j \in S_{0}} 1_{\{|w_j^{\prime} - \delta_j^{\prime} + \sqrt{n} \beta_j|>t\}}
\end{equation}
and $s_0/p = o(1)$, we see that the second summand in \eqref{eq5} converges almost surely to $0$ and that $\Var\{\bar R_{\mb{z}}\left(t\right)\} - \Var \{\bar{V}_{\mb{z}}(t) \} = \sop(1)$.
From \autoref{LmDiffExpectations} and \autoref{PropOrderCov}, we have
$\Var\{\bar V_{\mb{z}} \left(t\right) \} - \Var\{\bar V_{\tilde{\mb{w}}} \left(t\right) \} = \sop(1)$ and $\Var\{\bar V_{\tilde{\mb{w}}} \left(t\right) \} =\sop(1) $.
Therefore, $\Var\{\bar R_{\mb{z}}\left(t\right)\} = \sop(1)$. This completes the proof.

\subsection*{Proof of \autoref{Cor:FDPcontrol} }

First of all, the definitions of $t_{\alpha}$ and $\widehat{FDP}\left(  t_{\alpha}\right)$ imply%
\begin{equation} \label{eqq4}
\Pr\left\{  \widehat{FDP}\left(  t_{\alpha}\right)  \leq\alpha\right\}  =1
\end{equation}
and $\Pr\left\{2p \Phi(-t_\alpha) \le \alpha R_{\mb{z}}(t)\le \alpha p\right\} = 1$.
Then
\[
\Pr\left\{\Phi(-t_\alpha) \le \alpha/2\right\} = 1
\]
for a small constant $\alpha$, which implies that $t_\alpha$ does not go to $0$ as $p \to \infty$.
So, it suffices to consider positive constant values of $t_{\alpha}$.

Since the joint
distribution of $\left\{  z_{j}\right\}  _{j=1}^{p}$ and that of
$\{w_{j}^{\prime}\}_{j=1}^{p}$ remain the same conditional on $t_{\alpha}$,
identical arguments that led to \autoref{ThmSLLNFDP} and \autoref{Thm:mFDR} give%
\begin{equation}
\widehat{FDP}\left(  t_{\alpha}\right)  -FDP_{\mathbf{z}}\left(  t_{\alpha
}\right)  =o_{\Pr}\left(  1\right)  \text{ \ and \ }\widehat{FDP}\left(
t_{\alpha}\right)  -mFDR_{\mathbf{z}}\left(  t_{\alpha}\right)  =o_{\Pr}\left(
1\right)  ,\label{eqq1}%
\end{equation}
both conditional on $t_{\alpha}$. So, for any fixed constant
$a>0$,%
\begin{align}
& \lim_{p\rightarrow\infty}\Pr\left\{  \left\vert \widehat{FDP}\left(  t_{\alpha
}\right)  -FDP_{\mathbf{z}}\left(  t_{\alpha}\right)  \right\vert >a\right\}
\nonumber\\
& =\lim_{p\rightarrow\infty}\E\left\{  \E\left(  \left.  1_{\left\{  \left\vert
	\widehat{FDP}\left(  t_{\alpha}\right)  -FDP_{\mathbf{z}}\left(  t_{\alpha
	}\right)  \right\vert >a\right\}  }\right\vert t_{\alpha}\right)  \right\}
\nonumber\\
& =\E\left\{  \lim_{p\rightarrow\infty}\E\left(  \left.  1_{\left\{  \left\vert
	\widehat{FDP}\left(  t_{\alpha}\right)  -FDP_{\mathbf{z}}\left(  t_{\alpha
	}\right)  \right\vert >a\right\}  }\right\vert t_{\alpha}\right)  \right\}
\label{eqq2}\\
& =0\label{eqq3}%
\end{align}
where (\ref{eqq2}) follows from the dominated convergence theorem and
(\ref{eqq3}) from (\ref{eqq1}). Therefore, (\ref{eqq4}) and (\ref{eqq3})
together imply%
\[
\Pr\left\{  FDP_{\mathbf{z}}\left(  t_{\alpha}\right)  \leq\alpha\right\}
\rightarrow1.
\]
By almost identical arguments given above, we see%
\[
\lim_{p\rightarrow\infty}\Pr\left\{  \left\vert \widehat{FDP}\left(  t_{\alpha
}\right)  -mFDR_{\mathbf{z}}\left(  t_{\alpha}\right)  \right\vert >a\right\}
=0,
\]
which together with (\ref{eqq4}) implies $\Pr\left\{ mFDR_{\mathbf{z}}\left(
t_{\alpha}\right)  \leq\alpha\right\}  \rightarrow1$.

\bibliographystyle{myjmva}
%\bibliography{reference_FDP}

\end{document}